\documentclass[epj]{svjour}
\usepackage{graphicx}
\usepackage{amsmath}
\usepackage{amssymb}
\usepackage{amsfonts}
\usepackage{bm}
\def\be{\begin{equation}}
\def\ee{\end{equation}}
\def\bem{\begin{multline}}
\def\eem{\end{multline}}
\def\bea{\begin{eqnarray}}
\def\eea{\end{eqnarray}}
\def\RR{\mathbb{R}}

\begin{document}

\title{Internal structure of a quantum soliton and classical
excitations due to trap opening}

\author{Yvan Castin}
\institute{Laboratoire Kastler Brossel, \'Ecole Normale
Sup\'erieure, UPMC and CNRS, 24 rue Lhomond, 75231 Paris Cedex 05, France}

\date{Received: 14 July 2008 / Revised version: 19 September 2008}

\abstract{
We analytically 
solve two problems that may be useful in the context of the recent
observation of matter wave bright solitons in a one-dimensional attractive
atomic Bose gas. The first problem is strictly beyond mean field:
From the Bethe ansatz solution we extract the {\sl internal}
correlation function of
the particle positions in the quantum soliton, that is for a
{\sl fixed} center of mass position. 
The second problem is solved in the limit of a large number of
particles, where the mean field theory is asymptotically correct:
It deals with the number of excitations created by the opening of
the trap, starting from a pure soliton in a weakly curved
harmonic potential.
} 


\maketitle

Experiments with cold atoms have now acquired a high degree of
control of the key parameters of the system. Using transverse
confinement of the atoms by non-dissipative optical potentials
it is possible to freeze the atomic motion along one or several
directions, realizing in this way quantum gases with reduced
dimensionality \cite{low_dim}. Furthermore, thanks to Feshbach resonances
driven by a magnetic field, one can adjust almost at will the interaction
strength between the atoms \cite{feshbach_expt}.
The combination of these two experimental tools has recently
allowed the observation of bright solitons in a one-dimensional (1D)
Bose gas, either a single soliton 
\cite{Khaykovich} or a train of solitons \cite{Hulet}.

This leads to a renewed interest \cite{Mazets,Caux1,Caux2}
in the 1D non-relativistic Bose gas
on the free line
with zero range attractive interactions, that is with a binary
interaction potential modeled by a Dirac delta with a negative coupling
constant $g$, $V(x_1-x_2)=
g \delta(x_1-x_2)$, a model that is an acceptable approximation
of reality under conditions defined in \cite{Olshanii_g1d}.
On a theoretical point of view, it is well known that eigenstates
and eigenenergies of the corresponding $N$-body Hamiltonian may be
obtained by the Bethe ansatz:
Historically the focus was put mainly
on the study of the ground state wavefunction 
$\phi(x_1,\ldots,x_N)$ \cite{McGuire,Gaudin}, which
is a collective bound state of the $N$ particles, the so-called
$N$-particle quantum soliton, with a delocalized
center of mass of vanishing momentum.
The extension of the Bethe ansatz to excited states is however 
possible, and one finds that the generic excited 
state corresponds to a set of quantum solitons with arbitrary atom
numbers and different momenta per particle \cite{Herzog}.
A key property to keep in mind is the {\sl full separability} of the center
of mass variables and the internal variables of the gas 
(e.g. the relative coordinates of the particles), which holds 
since the gas is on the free line with open
boundary conditions.

Despite the knowledge of the ground state wavefunction for $N$ particles,
some theoretical work is needed to extract experimentally relevant
observables.
As atomic density profiles may be measured by absorption or even non-destructive
imaging \cite{Ketterle_probing}, natural observables are functions
of the positions of the particles.
The simplest observable is the mean density of particles,
$\rho(x)$, obtained by an average of the density profile over many
(ideally infinitely many) experimental realizations.
From the ground state $N$-body wavefunction one however does not obtain
any useful information on the mean density: 
Since the center of mass is fully delocalized,
one gets a uniform distribution over the whole line. 
Experimental reality is very different, the soliton being obtained
from an initially {\sl trapped} Bose-Einstein condensate.
When the trap is switched off to free the soliton, 
its center of mass is not in its ground state, it is in a localized
and non-stationary state which depends on the experimental
preparation procedure. A more realistic assumption is thus
to assume a $N$-body wavefunction of the form
\be
\Psi(x_1,\ldots,x_N) = \Phi(R) \phi(x_1,\ldots,x_N),
\label{eq:facto}
\ee
where 
\be
R=\frac{1}{N} \sum_{i=1}^{N} x_i
\ee
is the center of mass position of the soliton and the center of mass
wavefunction $\Phi(R)$ is {\sl a priori} unknown and depends
on the experimental details. The theoretical challenge is thus 
to predict observables for a {\sl fixed} position of the center of mass $R$.
Experimentally relevant results are then obtained from these theoretical
predictions by a further average over $R$ with the probability
distribution $|\Phi(R)|^2$.
One can even hope that the predictions for fixed $R$ are measurable,
if the center of mass position can be measured with a high enough
accuracy for each individual realization of the experiment.

Turning back to the simple example of the mean density, we see
that the right concept is the mean density of particles $\rho(x|R)$
for a fixed center of mass position $R$, as was already argued
in \cite{Herzog} within the general concept of (here translational)
symmetry breaking. Remarkably an explicit expression of $\rho(x|R)$
in terms of a sum of $N-1$ exponential terms may be obtained 
\cite{Calogero,Herzog_a_part}. 

The next step beyond the mean density is the pair distribution function 
of the particles $\rho(x,y)$, or very similarly the static structure
factor $S(x,y)$, of the $N$-particle soliton. 
Recently a general study of the dynamic structure factor was performed
from the Bethe ansatz \cite{Caux1,Caux2}, which includes the static
structure factor $S(x,y)$ and large $N$ expansions as limiting cases.
The goal of the present work is, in the spirit of the above physical 
discussion, to study the static structure factor $S(x,y|R)$
for a {\sl fixed} center of mass position $R$.
This static structure factor gives access to correlations between
the positions of the particles inside the soliton, that is it gives
information on the {\sl internal} structure of the soliton, which goes
beyond the usual mean field (or Gross-Pitaevskii) approximation, which
neglects such correlations.

As a guideline we imagine that, in an experiment, one wishes
to access the variance of a one-body observable of the gas
involving some function $U(x)$ of the particle position $x$:
\be
w(R) = \Big\langle \left[\sum_{i=1}^{N} U(x_i)\right]^2\Big\rangle_R
-\Big\langle \sum_{i=1}^{N} U(x_i)\Big\rangle_R^2,
\label{eq:def_w}
\ee
where the expectation value $\langle\ldots\rangle_R$ is taken
over the internal wavefunction $\phi(x_1,\ldots,x_N)$ of the quantum soliton
for a fixed center of mass location $R$.

The paper is organized as follows. In section \ref{sec:basic} 
we recall the basic facts about the 1D Bose gas model and we give
a general expression of the pair distribution function $\rho(x,y|R)$
for fixed $R$ that may be used to evaluate $w(R)$ numerically
for a moderate number of atoms and that will be the starting point
of analytical calculations.
In section \ref{sec:broad} we present an analytical calculation of $w(R)$
for an arbitrary atom number $N$, in the limit where the function
$U(x)$ is slowly varying over the spatial width of the soliton.
In section \ref{sec:asympt} we present a large $N$
expansion of the static structure factor $S(x,y|R)$ for fixed $R$
that we then use to calculate $w(R)$ to leading order in $N$;
in the same section, we integrate $S(x,y|R)$ over $R$
to see if we recover the results of \cite{Caux2} for $S(x,y)$,
and we also consider the case of a function $U(x)$
much narrower than the soliton.
In section \ref{sec:opening} we evaluate the accuracy of the
assumption (\ref{eq:facto}) in an experiment: Assuming that the
gas in the trap is in its ground state 
(at least for its internal variables),
we calculate analytically
(to leading order in $N$ and in the trap curvature) 
the number of internal excitations
of the soliton produced by the trap opening.
We conclude in section \ref{sec:conclusion}.

\section{Model, basic definitions and general results}
\label{sec:basic}

\subsection{Hamiltonian and ground state wavefunction}

We consider a set of $N\geq 2$ spinless non-relativistic bosons of mass
$m$ moving on the one dimensional real line, in the absence of any
trapping potential, with open boundary conditions, and binary interacting
{\sl via} an attractive contact potential of coupling constant $g<0$. 
This corresponds
to the Hamiltonian in first quantized form:
\be
H = \sum_{i=1}^{N} -\frac{\hbar^2}{2m}\, \partial_{x_i}^2 
+ \sum_{1\leq i < j \leq N} g \delta(x_i-x_j).
\label{eq:hamil}
\ee
For this problem there is full separability of the center of mass motion
and of the internal coordinates, the internal wavefunction being independent
of the center of mass state. As a consequence, the ground state wavefunction
has it center of mass with zero momentum and depends only on the relative
coordinates of the particles. Its exact expression is \cite{McGuire}
\be
\phi(x_1,\ldots,x_N) = \mathcal{N} \exp
\left[
-\frac{m|g|}{2\hbar^2} \sum_{1\leq i < j \leq N} |x_i-x_j|
\right]
\ee
where the normalization factor $\mathcal{N}$ will be specified later. 
The corresponding eigenenergy is \cite{McGuire}
\be
E_0(N) = -\frac{mg^2}{24\hbar^2} (N-1)N(N+1).
\ee
This ground state is the so-called quantum soliton with $N$ particles.

\subsection{Mean density, pair distribution and static structure factor
for a fixed center of mass position}

The crucial concept in the present work is the expectation
value of a position dependent observable, that is of an arbitrary
function $O(x_1,\ldots,x_N)$ of the $N$ particle positions, for a {\sl fixed}
value $R$ of the center of mass position:
\bea
\langle O\rangle_R &\equiv&
\int dx_1\ldots dx_N \, \delta\left(R-\frac{1}{N}\sum_{k=1}^{N} x_k\right) 
\nonumber \\
&& \times O(x_1,\ldots, x_N) |\phi(x_1,\ldots,x_N)|^2,
\label{eq:def_moy}
\eea
where the integration is over the whole space $\RR^N$.
The soliton wavefunction shall then be normalized in such a way that
the expectation value $\langle 1\rangle_R$ of the function $O$ constant and
equal to unity is also equal to unity. 
Using the bosonic symmetry of the wavefunction, the integral to compute
is then simply $N!$ times the integral over the so-called {\sl fundamental} domain 
of ordered positions
\be
D={(x_1,\ldots,x_N)\ \ \mbox{such that}\ \  x_1 < \ldots < x_N}.
\label{eq:fd}
\ee
Over this domain, the wavefunction indeed has the separable expression
\be
\phi(x_1,\ldots,x_N) = \mathcal{N} e^{-\frac{m|g|}{2\hbar^2} \sum_{j=1}^{N}
[2j-(N+1)]x_j}.
\label{eq:phi}
\ee
With the change of variables (\ref{eq:cov}) 
one then obtains
\be
|\mathcal{N}|^2 \left(\frac{\hbar^2}{m|g|}\right)^{N-1}\frac{N}{(N-1)!} =1.
\ee

Setting $O=\sum_j \delta(x-x_j)$ we obtain the mean density $\rho(x|R)$ 
for a fixed center of mass position. This was first calculated in
\cite{Calogero}:
\be
\rho(x|R) = \frac{N!^2}{N\xi} \sum_{k=0}^{N-2} \frac{(-1)^k(k+1)}{(N-2-k)!(N+k)!}
e^{-(k+1)|x-R|/\xi},
\label{eq:calogero}
\ee
where $\xi$ is the spatial width of the classical (that is mean-field) soliton,
\be
\xi = \frac{\hbar^2}{m |g| N}.
\label{eq:xi}
\ee
It is a function of $|x-R|$, a consequence of translational and parity
invariance of the Hamiltonian and of $\phi$. 
One thus has $\rho(x|R)=\rho(x-R|0)=\rho(R-x|0)$.
It is normalized in such a way that the integral over $x$
over the whole space is equal to $N$.
If the true physical state of the system is a product 
(\ref{eq:facto}) of a localized center of mass wavefunction $\Phi(R)$ 
and of the quantum soliton wavefunction, the physical mean density
is the average (or equivalently the convolution)
\be
\rho(x) = \int_{-\infty}^{+\infty} dR\, |\Phi(R)|^2 \rho(x-R|0).
\ee

Setting $O=\sum_{i\neq j} \delta(x-x_i) \delta(y-x_j)$, we obtain
the pair distribution function $\rho(x,y|R)$ for fixed center of mass
position $R$. It is normalized as $N(N-1)$ for the double integration
over $x$ and $y$. From the translation invariance
$\rho(x,y|R)=\rho(x-R,y-R|0)$ so that it is sufficient to
calculate this pair distribution function for $R=0$.
In real space, we do not know an expression of $\rho(x,y|0)$
as simple as (\ref{eq:calogero})
but we have found a simple expression for the Fourier transform
\be
\tilde{\rho}(q_a,q_b|0) \equiv
\int_{\RR^2} dx\, dy\, \rho(x,y|0) e^{-i(q_a x + q_b y)}.
\ee
As detailed in the appendix \ref{appen:deriv_gen} one has
\begin{multline}
\tilde{\rho}(q_a,q_b|0) = \left|\frac{\Gamma(N)}{\Gamma(N+iQ)}\right|^2\, 
\sum_{1\leq j< k \leq N} \\
\frac{\Gamma(k-iQ)\Gamma(N+1+iQ-j)}{\Gamma(j)\Gamma(N+1-k)} \\
\times \left[ 
\frac{\Gamma(j-\epsilon_a)\Gamma(N+1+iQ-\epsilon_a-k)}
{\Gamma(k-\epsilon_a)\Gamma(N+1+iQ-\epsilon_a-j)}
+ \epsilon_a\leftrightarrow \epsilon_b\right]
\label{eq:paire_fs}
\end{multline}
where $\Gamma(z)$ is the Gamma function of complex argument $z$ and
we have introduced the dimensionless variables
$Q_{a,b} = q_{a,b}\xi$, 
$Q = Q_a+Q_b$ and $q=(Q_b-Q_a)/2$.
The quantities $\epsilon_{a,b}$ solve the degree two equations:
\be
\epsilon_{a,b} (N+iQ-\epsilon_{a,b}) = i Q_{a,b} N.
\label{eq:trin}
\ee
In the large $N$ limit, for fixed values of $Q_a$ and $Q_b$, it is convenient
to take
\be
\epsilon_{a,b}  = \frac{1}{2}(N+iQ) -[(iq\pm N/2)^2-Q_aQ_b]^{1/2}
\label{eq:sol_trin}
\ee
with a determination of the square root 
such that $\epsilon_{a,b}= i Q_{a,b} +Q_a Q_b/N + O(1/N^2)$ for $N\to +\infty$,
e.g.\ with the line cut on the real negative axis.

This gives the idea {\sl a posteriori} to look for a similar expression 
for the Fourier transform $\tilde{\rho}(q_a|0)$ of $\rho(x|0)$.
As shown in the appendix \ref{appen:deriv_gen} one obtains
the simple expression
\begin{multline}
\tilde{\rho}(q_a|0) = 
 \left|\frac{\Gamma(N)}{\Gamma(N+iQ_a)}\right|^2\, \\
\times \sum_{j=1}^{N} 
\frac{\Gamma(j-iQ_a) \Gamma(N+1+iQ_a-j)}{\Gamma(j) \Gamma(N+1-j)}.
\label{eq:rho_fs}
\end{multline}
Amusingly, this allows to express $\tilde{\rho}(q_a|0)$ in
terms of the standard hypergeometric function evaluated in $z=1$,
\begin{multline}
\tilde{\rho}(q_a|0) = \frac{\Gamma(N)\Gamma(1-iQ_a)}{\Gamma(N-iQ_a)} \times\\
{}_2 F_1(1-N,1-iQ_a;1-N-iQ_a;1),
\end{multline}
although we have not found this expression particularly useful.

Let us come back to the original problem of calculating 
the variance $w(R)$ of the one-body observable $\sum_i U(x_i)$, 
as defined in (\ref{eq:def_w}). One first expresses the squares as products of
double sums over indices $i$ and $j$. One can split the double sum over $i$ and $j$
in (\ref{eq:def_w}) in diagonal terms $i=j$ and off-diagonal terms $i\neq j$,
so that $w(R)=w_{\rm diag}(R) + w_{\rm off}(R)$.
The diagonal terms can be expressed in terms of the mean density,
\begin{multline}
w_{\rm diag}(R) \equiv 
\sum_{i=1}^N \langle U(x_i)^2\rangle_R -\left[\sum_{j=1}^N
\langle U(x_j)\rangle_R\right]^2 \\
= \int_{\RR} dx \rho(x|R) U(x)^2 -\left[\int_{\RR} dx \rho(x|R) U(x)\right]^2.
\end{multline}
This can be directly evaluated from (\ref{eq:calogero}).
The off-diagonal terms depend on the pair distribution function,
\begin{multline}
w_{\rm off}(R) = \sum_{i\neq j} \langle U(x_i) U(x_j)\rangle_R \\
= \int_{\RR^2} dx\, dy\, U(x) U(y) \rho(x,y|R). 
\end{multline}
Introducing the Fourier transform of $U(x)$,
\be
\tilde{U}(q_a)=\int_{\RR} dx U(x) \exp(-iqx),
\ee
we obtain the Fourier space expression
\begin{multline}
w_{\rm off}(R) 
= \int_{\RR^2} \frac{dq_a dq_b}{(2\pi)^2} \tilde{U}(q_a) \tilde{U}(q_b)
\\ \times 
e^{i(q_a+q_b)R} \tilde{\rho}(-q_a,-q_b|0)
\label{eq:w_off}
\end{multline}
that can be directly
evaluated using (\ref{eq:paire_fs}).
Simpler analytic expressions of $w(R)$, either for slowly varying
functions $U(x)$ or in the large $N$ limit, shall be given in the 
next sections.

To conclude this subsection, we note that $w(R)$ has a very
simple expression in terms of the correlations contained in
the fixed $R$ static structure factor $S(x,y|R)$,
\begin{multline}
S(x,y|R) \equiv \langle \hat{\rho}(x) \hat{\rho}(y)\rangle_R \\
=  \delta(x-y) \rho(x|R) + \rho(x,y|R)
\label{eq:Svsrho}
\end{multline}
where the operator giving the density is
\be
\hat{\rho}(x) = \sum_{i=1}^{N} \delta(x_i-x).
\ee
We note that the double integral of $S(x,y)$ over $x$ and $y$ is
equal to $N^2$.
Using again the translational invariance, one obtains the illuminating
expression
\begin{multline}
w(R) = \int_{\RR^2} dx\, dy\,
U(x+R) U(y+R) \\
\times \left[S(x,y|0)-\rho(x|0)\rho(y|0)\right].
\label{eq:illumi}
\end{multline}
The essence of the mean-field approximation is to neglect correlations
among the particles, so that $S(x,y|0)$ would essentially be approximated
by the uncorrelated product $\rho(x|0)\rho(y|0)$. The above writing
clearly reveals that $w(R)$ is sensitive to correlations that are beyond
the mean-field approximation.

\subsection{The usual static structure factor}

The usual static structure factor is the spatial correlation function 
of the operator
$\hat{\rho}(x)$ giving the density, 
\be
S(x,y) = \langle \hat{\rho}(x) \hat{\rho}(y)\rangle
\ee
where the expectation value is taken literally in the ground state of the gas,
thus assuming a perfectly delocalized center of mass wavefunction
$\Phi(R)=1$. Because of the translational invariance, it is a function
of $x-y$ only.

This usual structure factor is deduced from our fixed-$R$ one by integration 
over the center of mass position
\be
S(x,y) = \int_{\RR} dR\, S(x,y|R) = \int_{\RR} dR\, S(x-R,y-R|0).
\label{eq:link}
\ee
This allows to express 
the Fourier transform of $S(x,y)$ in terms of the Fourier transform
of $\rho(x,y|0)$, when one uses (\ref{eq:Svsrho}):
\be
\tilde{S}(q_a,q_b)=2\pi\delta(q_a+q_b)\left[N+\tilde{\rho}(q_a,q_b|0)\right].
\ee
From (\ref{eq:paire_fs}) we thus have an analytical expression of 
$\tilde{S}(q_a,q_b)$ in terms of a double sum.

The relation (\ref{eq:link})  will also allow us, 
in the large $N$ limit, to convert 
our large $N$ expansion of $S(x,y|R)$ into
a large $N$ expansion of $S(x,y)$, see \S\ref{subsec:link}. 

\section{Value of $w(R)$ for a broad function $U(x)$}
\label{sec:broad}

One supposes in this section that $U(x)$ varies slowly over the length scale
$\xi$ of the quantum soliton, as defined in (\ref{eq:xi}). 
Then one rewrites (\ref{eq:def_w}) using the translational invariance,
\be
w(R) = \langle \left[\sum_{i=1}^{N} U(x_i+R)\right]^2\rangle_0
-\langle \sum_{i=1}^{N} U(x_i+R)\rangle_0^2,
\ee
and one expands
\be
U(x_i+R) = U(R) + x_i U'(R) + \frac{1}{2} x_i^2 U''(R)+\ldots
\ee
The constant shift $U(R)$ has no effect in $w(R)$. The linear term has also
an exactly vanishing contribution, since by definition
$\langle (\sum_{i=1}^{N} x_i)^n\rangle_0 = 0$ for all integers
$n\geq 1$.
Setting
\be
O_2 = \sum_{i=1}^{N} x_i^2,
\label{eq:defO2}
\ee
we thus obtain the leading contribution
\be
w(R) \simeq \frac{1}{4} \left[U''(R)\right]^2 \left(\mbox{Var}\,O_2\right)_0,
\ee
with $\left(\mbox{Var}\,O_2\right)_0=\langle O_2^2\rangle_0-\langle O_2\rangle_0^2$
is the variance of $O_2$ for a center of mass position fixed at the origin
of the coordinates.

It turns out that an exact expression may be obtained for this variance,
as detailed in the appendix \ref{appen:deriv_broad}:
It is the sum of three contributions,
\be
\left(\mbox{Var}\, O_2\right)_0 
= \left(\frac{\hbar^2}{m|g|}\right)^4 \left[S_1 + S_2 + S_3\right],
\label{eq:sotc}
\ee
with
\bea
S_1 &=& 4 \sum_{i=2}^{N} \sum_{j=2}^{N} \sum_{k=2}^{N} B_{ij}B_{ik} \\
S_2 &=& \sum_{i=2}^{N} \sum_{j=2}^{N} \left[8 B_{ii} B_{ij}+2 B_{ij}^2\right] \\
S_3 &=&  6 \sum_{i=2}^{N} B_{ii}^2.
\eea
We have introduced the symmetric matrix
\be
B_{ij}= \frac{1}{N} \frac{1}{[N+1-\mbox{min}(i,j)][\mbox{max}(i,j)-1]},
\label{eq:defB}
\ee
defined over the index range $2\leq i,j\leq N$. This holds whatever the value
of the atom number $N\geq 2$.

In the large $N$ limit, the contribution $S_1$ is dominant, simply because
it contains more terms, and its asymptotic expression is evaluated by
replacing the sums by integrals,
\begin{multline}
S_1 \simeq \frac{4}{N^3} \int_0^{1} dx\, \left[\frac{\ln(1-x)}{x}+\frac{\ln x}{1-x}\right]^2 \\ 
= N^{-3}\left[\frac{8\pi^2}{3} + 16 \zeta(3)\right].
\end{multline}
This leads to the estimate
\be
w(R) \simeq N\xi^4 \left[U''(R)\right]^2 \left[\frac{2\pi^2}{3} +4\zeta(3)\right],
\label{eq:wslow}
\ee
for a slowly varying potential $U(x)$ in the $N\gg 1$ limit.

\section{Large $N$ limit of $w(R)$ for $U(x)$ of any width}
\label{sec:asympt}

In this section we give a large $N$ expansion of the static structure factor
for fixed center of mass position, which allows to get the asymptotic behavior
of $w(R)$ in the large $N$ limit. When specialized to a quadratic potential
$U(x)$ the general result reproduces the large $N$ broad potential
result of the previous section.
As a first test of the result,  
we integrate $S(x,y|R)$  over the center of mass 
position $R$, to see if we recover 
the results of \cite{Caux2} for the usual static factor $S(x,y)$.
As a second test of the result, we get an approximate expression 
for $w(R)$ in the case of a narrow function $U(x)$, to lowest
order in the width $b$ of $U(x)$, first from the large
$N$ expansion of $S(x,y|R)$ and then from a more general reasoning
not relying on a large $N$ expansion.

\subsection{Asymptotic expression of $S(x,y|R)$ and of $w(R)$}

As we have seen in (\ref{eq:illumi}), $w(R)$ is directly related to the deviation of
the fixed $R$ static structure factor of the soliton from the uncorrelated form
$\rho(x|R)\rho(y|R)$.
It turns out that this deviation may be easily obtained from the Fourier
space expressions (\ref{eq:paire_fs},\ref{eq:rho_fs}), 
simply by taking the large $N$ limit
of the $\Gamma$ functions and by replacing the discrete sums over indices
by integrals. This shows that our Fourier space representations are indeed
useful.

As detailed in the appendix \ref{appen:asympt}, the large $N$ expansion
for a fixed value of $\xi$ (that is for a fixed value of $N g$),
gives the leading term
\begin{multline}
S(x,y|0)-\rho(x|0)\rho(y|0) \simeq 
-N \partial_x \partial_y  \\
\left[\theta(y-x)\frac{y/\xi-x/\xi-e^{x/\xi}-e^{-y/\xi}}
{[2\cosh(x/2\xi)]^2[2\cosh(y/2\xi)]^2}+x\leftrightarrow y\right]
\label{eq:Sasympt}
\end{multline}
where $\theta(x)$ is the Heaviside distribution that is equal to zero for $x<0$ and
to one for $x>0$.
If one wishes to have an expansion of the static structure factor only, one has
also to expand $\rho(x|0)$ in powers of $N$. 
From the real space expression (\ref{eq:calogero}), expanding
each factorial in the large $N$ limit for a fixed summation index $k$,
we obtain, setting $X=x/\xi$, a result in agreement with \cite{Calogero}:
\be
\rho(x|0) = \frac{N}{\xi} \left[1-\frac{1}{N} \frac{d^2}{dX^2}+\ldots\right]
\frac{1}{[2\cosh(X/2)]^2},
\label{eq:rho_exp}
\ee
where one may check the normalization condition $N=\int_{\RR} dx\, \rho(x|0)$.
This gives the expansion of $S(x,y|0)$ up to order $N$, for a fixed value of $\xi$.

\begin{figure}
\includegraphics[width=8cm,clip]{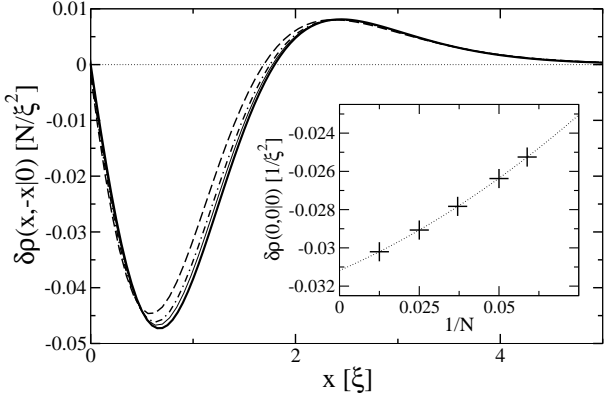}
\caption{Cut along the line $y=-x$ of the function
$\delta\rho(x,y|0)=\rho(x,y|0)-\rho(x|0)\rho(y|0)$.
Thick solid line: Large $N$ limit (\ref{eq:drho_asympt}). Dashed line:
Numerical result for $N=10$. Dot-dashed line: Numerical result for $N=20$.
Thin solid line: Numerical result for $N=40$.
$\delta\rho$ is in units of $N/\xi^2$, and the coordinate
$x$ is in units of $\xi$. The inset shows numerical evidence
for the convergence of $\delta\rho(0,0|0)\xi^2$ towards a non-zero value
$\simeq -0.031$ at large $N$,  
which is compatible with the analytical prediction $\delta\rho(0,0|0)\xi^2/N
\to 0$. In the inset, symbols are numerical data,
and the dotted line is a quadratic
fit used to guide the eye.
}
\label{fig:drho}
\end{figure}

From (\ref{eq:Svsrho}) and (\ref{eq:Sasympt}) we can directly
obtain, in the large $N$ limit, the pair correlations between the positions
of the particles for a fixed center of mass position:
\be
\delta\rho(x,y|R)\equiv \rho(x,y|R)-\rho(x|R)\rho(y|R). 
\ee
One can indeed show that the distributions
generated in (\ref{eq:Sasympt}) by the derivatives of 
$\theta(x-y)$ and $\theta(y-x)$ with respect
to $x$ and $y$ exactly cancel with the Dirac term appearing
in (\ref{eq:Svsrho}). As a consequence the expression for
$\delta\rho(x,y|0)$ is deduced from the right hand side of (\ref{eq:Sasympt}) 
simply by exchanging the order of the $\theta$ distributions  and of the operator
$\partial_x\partial_y$.
After an explicit calculation of the derivatives with respect to $x$ and $y$
we obtain
\begin{multline}
\delta\rho(x,y|0)\simeq -\frac{N}{16\xi^2} 
\left[(2+|X-Y|)\sinh(X/2)\sinh(Y/2) \right. \\ \left.+2\sinh(|X-Y|/2)\right]/
\left[\cosh(X/2)\cosh(Y/2)\right]^3,
\label{eq:drho_asympt}
\end{multline}
with $X=x/\xi$ and $Y=y/\xi$. A contour plot of this function reveals
that it has an interesting structure, in the form of two valleys
separated by a crest on the $x=y$ line, each valley
being elongated in the direction parallel to $x=y$ and
containing two local minima separated by a saddle point. 
For clarity we only show
a plot of $\delta\rho$ along the line $y=-x$, restricting to
$x>0$ by parity, see the thick solid
line in Fig.\ref{fig:drho};
the minimum of this line
then corresponds in the full $x-y$ plane
to one of the aforementioned saddle points.
In the same figure, we also give the value of $\delta\rho$
for finite values of $N$, obtained by calculating the Fourier 
transform of (\ref{eq:fn1}) and (\ref{eq:fn2}) numerically.
This shows that the large $N$ limit is well approached with moderately
high values of $N$ already.

We come back to (\ref{eq:Sasympt}) to obtain the large $N$ equivalent of $w(R)$
at fixed $\xi$.
By repeated integration by parts in the double integral
of (\ref{eq:illumi}), and using the fact that
\be
-\frac{Y-X-e^X-e^{-Y}}{[2\cosh(X/2)]^2[2\cosh(Y/2)]^2} =
\partial_X \partial_Y \frac{2+Y-X}{(e^Y+1)(e^{-X}+1)},
\label{eq:primi}
\ee
we find \cite{aide_sup} for a function $U(x)$ not rapidly increasing at 
$|x|=\infty$ (that is not increasing faster than a power law):
\begin{multline}
w(R) \simeq 2N\xi^4  \int_{-\infty}^{+\infty}dX\, \int_{X}^{+\infty} dY\,
\\
U''(R+X\xi) U''(R+Y\xi) \frac{2+Y-X}{(e^Y+1)(e^{-X}+1)}.
\label{eq:wgn}
\end{multline}

One immediately sees that  a linearly varying potential $U(x)$ 
gives a vanishes contribution to $w(R)$, which is also obvious
from the definition (\ref{eq:def_w}), 
since $\langle (\sum_i x_i)^n\rangle_R =(N R)^n$
by construction.
If $U(x)$ is a quadratic function of $x$, $U(x)=U''(R) x^2/2$,
one should recover Eq.(\ref{eq:wslow}), which is indeed
the case if
\be
\frac{\pi^2}{3}+2\zeta(3) = \int_{-\infty}^{+\infty}dX\, \int_{X}^{+\infty} dY\,
\frac{2+Y-X}{(e^Y+1)(e^{-X}+1)}.
\ee
We have checked that this identity holds
\cite{details}.

\subsection{Application: large $N$ expansion of the usual static structure factor}
\label{subsec:link}

The usual static structure factor $S(x,y)$ is obtained from
$S(x,y|R)$ by integration over $R$, see (\ref{eq:link}).
Assuming without loss of generality that $y=0$, we see from (\ref{eq:Sasympt}) 
and from (\ref{eq:primi}) that one has to integrate over $R$ quantities
of the form $(\partial_x \partial_y F)(x-R,R)$, where the function 
$F$ corresponds in a first stage to the expression in between square 
brackets in (\ref{eq:Sasympt}) and in a second stage to the function
over which $\partial_X \partial_Y$ acts in the right
hand side of (\ref{eq:primi}).
From the differential relations, taking $x$ and $R$ as independent variables,
\bea
(\partial_x F)(x-R,-R) &=& \frac{d}{dx} [F(x-R,-R)] \\
(\partial_y F)(x-R,-R) &=& -\left(\frac{d}{dx} + \frac{d}{dR}\right) [F(x-R,-R)], \ \ \  \ \ \ 
\eea
we see that the integral over $R$ will cancel the derivatives $d/dR$.
Using 
\be
\int_{-\infty}^{+\infty} dR \frac{1}{(e^{R}+1)(e^{-R+X}+1)}= \frac{X}{e^{X}-1},
\ee
one gets
\begin{multline}
S(x,0) -\int_{-\infty}^{+\infty} dR\, \rho(x|R) \rho(0|R) \simeq 
\\ \frac{N}{\xi} \frac{d^2}{dX^2}
\left[\theta(X)\frac{d^2}{dX^2}\frac{X(X+2)}{e^{X}-1}
+X\leftrightarrow -X\right],
\end{multline}
where $X=x/\xi$. One can show that the $\theta(X)$ distribution can be exchanged
with the last $d^2/dX^2$ if one wishes, but we shall not use this property
here.
From the large $N$ expansion (\ref{eq:rho_exp}) of $\rho(x|R)$, 
and using the fact that the second order derivative of $X/[\exp(X)-1]$
is an even function,
one finally obtains the large $N$ expansion
\begin{multline}
S(x,y=0) = \frac{N^2}{\xi} \frac{d^2}{dX^2}\left[ \frac{X}{e^X-1}\right] \\
+\frac{N}{\xi} \frac{d^2}{dX^2}\left[\theta(X)\frac{d^2}{dX^2}\frac{X^2}{e^X-1}+
X\leftrightarrow -X\right]+\ldots...
\label{eq:Soasymp}
\end{multline}
To compare this result with the ones of \cite{Caux2} given in Fourier space,
one performs the Fourier transform of Eq.(62) and Eq.(63) of \cite{Caux2}.
Then one finds that (\ref{eq:Soasymp})
is consistent with \cite{Caux2} if one adds in Eq.(62) of \cite{Caux2}
the subleading term of the large $N$ expansion of the
intermediate quantity $S_N^\rho(k)$ introduced in \cite{Caux2}. 

Furthermore, (\ref{eq:Soasymp}) can be shown to be equivalent to the 
more appealing writing, where the expected Dirac delta contribution 
is singled
out,
\begin{multline}
S(x,y=0)= \frac{N^2}{\xi} \frac{d^2}{dX^2} 
\left[\frac{X/2}{\tanh(X/2)}\right]
+N \delta(x) \\ +
\frac{N}{\xi} \left\{\theta(X)\frac{d^4}{dX^4}\left[\frac{X^2/2}
{\tanh(X/2)}\right]+X\leftrightarrow -X\right\}
+\ldots
\end{multline}
where we recall that $X=x/\xi$.
One can also easily check that this obeys the sum rule $\int_{\RR} dx\, S(x,y=0)=N^2$.

\subsection{Approximation of $w(R)$ for a narrow barrier}

In this subsection, one assumes that $U(x)$ and its derivatives are functions 
localized around the origin
with a width much smaller than $\xi$. E.g. $U(x)$
is a Gaussian centered in $x=0$ with a width $\ll \xi$.
One first considers the large $N$ limit.  We rewrite Eq.(\ref{eq:wgn}) as
\begin{multline}
w(R) \simeq  2N\xi^2 \int_{-\infty}^{+\infty} dx \int_x^{+\infty} dy\,
\\ U''(x) U''(y)  F\left(\frac{x-R}{\xi},\frac{y-R}{\xi}\right)
\end{multline}
with
\be
F(X,Y) = \frac{2+Y-X}{(e^{Y}+1)(e^{-X}+1)}.
\ee
Then we expand the factor containing $F$ in powers of $x/\xi$ and $y/\xi$.
If $U(x)$ and its derivatives are rapidly decreasing functions,
one can show that the first non vanishing contribution to $w(R)$
comes from the third order expansion, which results in
\be
w(R) \simeq \rho(R|0) \int_{-\infty}^{+\infty} dx\, U(x)^2,
\label{eq:case_narrow}
\ee
where we have used (\ref{eq:rho_exp}) to leading order in $N$ to recognize 
the factor $\rho(R|0)$.

There is actually a faster way to obtain this result, and {\sl not} restricted
to the large $N$ limit, from the exact writing
\begin{multline}
w(R) = \int dx\, \rho(x|R) U(x)^2 + \int_{\RR^2} dx\, dy\, \\
[\rho(x,y|R)-\rho(x|R)\rho(y|R)] U(x)U(y).
\label{eq:wvsrho}
\end{multline}
Then one sees that the first term in (\ref{eq:wvsrho})
is first order in the width $b$ of $U(x)$,
since the integration range over $x$ has a width $b$,
whereas the second term is second order in $b$, since it involves 
a double integral over a range of diameter $\sim b$. 
Then to first order in $b$ one recovers (\ref{eq:case_narrow}).
In the large $N$ limit, for a fixed $\xi$,
both terms of (\ref{eq:wvsrho}) scale in the same way with $N$, 
that is linearly with $N$, so that the
order of the limits $N\to +\infty$ and $b\to 0$ (for fixed $\xi$) 
is not crucial.

\section{Number of internal excitations created by the trap opening}
\label{sec:opening}

In this work, we have assumed up to now that a pure soliton is produced
in the experiment, corresponding to the state (\ref{eq:facto}): 
The center of mass may be in an arbitrary
excited state but the internal variables of the gas are in their ground state.
In this section, we revisit this assumption taking into account
experimental constraints.

In a real experiment the ultracold gas is prepared in a trap; it is not free
along $x$, and each atom is subject to the harmonic confining potential
\be
W(x) = \frac{1}{2} m \omega^2 x^2.
\label{eq:trapW}
\ee
We recall that the center of mass motion and the internal variables
remain separable in a harmonic trap.
In order for the trapped gas to be close to the free space
limit, the oscillation frequency $\omega$ is adjusted to have
$\hbar \omega \ll |\mu_0|$,
where
\be
\mu_0 = -\frac{\hbar^2}{8m\xi^2}
\ee
is the mean-field approximation for the chemical potential of the free-space soliton.
We assume that the trapped gas is cooled down to a temperature $T$ low enough
to have $k_B T \ll |\mu_0|$. The gas is then a pure soliton.
On the contrary we do not assume the much more stringent condition
$k_B T \ll \hbar\omega$, so that the center of mass of the gas may 
still be in an excited state.

Eventually the trapping potential along $x$ will be swit\-ched off,
to obtain the ideal conditions of the Hamiltonian
(\ref{eq:hamil}).
This trap opening will create some internal excitations of the gas,
that is the untrapped gas will not be a pure soliton but will
contain a mean number $N_{\rm exc}$ of internal excitations.

Since no Bethe ansatz solution exists in a trap, the modest goal
here is to calculate $N_{\rm exc}$ to leading order in $N$
in the large $N$ limit, for a fixed value of $|\mu_0|/(\hbar \omega) \gg 1$
\cite{lien}.
In this limit, a 1D classical field treatment is sufficient, 
with a Hamiltonian
\be
\mathcal{H}= \int_{\RR} dx\, \left[-\frac{\hbar^2}{2m} 
\psi^*\partial_x^2\psi+\frac{g}{2} \psi^{*2}\psi^2 +
\frac{1}{2} m\omega^2(t) x^2 \psi^* \psi\right].
\ee
We have slightly generalized (\ref{eq:trapW}) to treat the case
of a time dependent trap:
\be
\omega^2(t) = \omega^2 \chi(t)
\ee
where the function $\chi(t)$, going from unity for $t=0$ to zero for 
$t\to +\infty$ describes the switch-off procedure of the trap.
The atom number is fixed so that 
the norm squared of the classical field is also fixed:
\be
\int_{\RR} dx\, |\psi(x)|^2 = N.
\label{eq:nat}
\ee
This classical field problem can be solved on a computer, which will allow
a test of our predictions for $N_{\rm exc}/N$.

The problem can be further simplified in the limit $\hbar\omega/|\mu_0|\to 0$,
where one expects that the field $\psi(x)$ will remain ``close" 
(up to a phase factor)
to the one describing a pure soliton \cite{mean-field-soliton},
$\psi_0(x) = N^{1/2} \phi_0(x)$,
with
\be
\phi_0(x) = \frac{1}{2\xi^{1/2}} \frac{1}{\cosh[x/(2\xi)]}.
\ee
We then use the number conserving Bogoliubov formalism of 
\cite{Gardiner,CastinDum}, 
downgraded to a classical field problem (simply replacing
commutators with Poisson brackets), as was already done
in \cite{Emilia}. One splits the field by projection 
along the mode $\phi_0$ and orthogonally to it:
\be
\psi(x) = a_0 \phi_0(x) +\psi_\perp(x)
\label{eq:splitting}
\ee
where $a_0$ is the component of the field on the mode $\phi_0$
and the field $\psi_\perp(x)$ is orthogonal to that mode.
The idea is to treat $\psi_\perp(x)$ as a small perturbation.
The strength of the number conserving approach is to eliminate 
the amplitude $a_0$ in a systematic way, using the modulus-phase 
representation
\be
a_0 = |a_0| e^{i\theta}.
\ee
The phase $\theta$ is eliminated by a redefinition of the transverse field:
\be
\Lambda(x) \equiv e^{-i\theta} \psi_\perp(x).
\ee
The modulus $|a_0|$ is expressed in terms of $\Lambda$ using the 
condition of a fixed atom number (\ref{eq:nat}).

In the absence of trapping potential, one keeps terms up to quadratic
in $\Lambda$ in the Hamiltonian, and the resulting quadratic form 
can be written in normal form as \cite{Olshanii_u}
\be
\mathcal{H}_0 \simeq \mathcal{E}_0 + \frac{P^2}{2mN} 
+\int_{\RR} \frac{dk}{2\pi} \epsilon_k b_k^* b_k
\label{eq:normal}
\ee
where $\mathcal{E}_0=N\mu_0/3$ is the ground state of the 
classical field model. The variable $P$ of the field represents
the total momentum of the field, written to first order in $\Lambda$:
\be
P = \frac{\hbar}{i} N^{1/2} \int_{\RR} dx\ \phi_0'(x) [\Lambda^*(x)-\Lambda(x)].
\ee
The occurrence of the term $P^2/(2mN)$ represents physically the fact that the
center of mass motion is decoupled; more formally, it corresponds
to the fact that $\phi_0(x)$ ``spontaneously" breaks the translational
symmetry of the Hamiltonian, which leads to the occurrence of a Goldstone
mode \cite{Ripka,Lewen,Ueda}. The field variable canonically conjugated to $P$
corresponds to the center of mass position of the field, written
up to first order in $\Lambda$:
\be
Q = N^{-1/2} \int_{\RR} dx\, x \phi_0(x) [\Lambda(x)+\Lambda^*(x)].
\ee
Apart from this Goldstone mode, the other eigenmodes behave as
a continuum of decoupled harmonic oscillators, with normal (complex) variables
$b_k$ and eigenenergies
\be
\epsilon_k = |\mu_0| + \frac{\hbar^2 k^2}{2m},
\ee
which correspond to internal (and thus gapped) 
excitations of the gas: An elementary excitation
physically takes the form of a free particle coming from infinity with a
wavevector $k$ and scattering on a soliton with $N-1$ particles.
The field variable $b_k$ has the expression
\be
b_k \equiv \int_{\RR} dx\ \left[u_k^*(x) \Lambda(x) -v_k^*(x) \Lambda^*(x)
\right]
\ee
where the Bogoliubov modes of the number conserving theory are
expressed as follows in Dirac's notation 
(see \cite{CastinDum} \S V.A)  here for a real function $\phi_0$:
\bea
|u_k\rangle &=& \mathcal{Q} |U_k\rangle \\
|v_k\rangle &=& \mathcal{Q} |V_k\rangle
\eea
where $\mathcal{Q}=1-|\phi_0\rangle\langle\phi_0|$ projects orthogonally
to $|\phi_0\rangle$, and the $U_k,V_k$ are the eigenmodes of the usual
Bogoliubov-de Gennes equations
\bea
\epsilon_k U_k(x) & =& 
\left[-\frac{\hbar^2}{2m}\partial_x^2 +2 g N |\phi_0(x)|^2-\mu_0\right]U_k(x)
\nonumber \\
&&+ g N \phi_0^2(x) V_k(x) \\
-\epsilon_k V_k(x) & =& \left[-\frac{\hbar^2}{2m}\partial_x^2 +2 g N |\phi_0(x)|^2-\mu_0\right]V_k(x)
\nonumber \\
&& + g N \phi_0^{*2}(x) U_k(x).
\eea
It turns out that these modes are known exactly \cite{Kaup}. For
$k>0$ one has
\bea
U_k(x) &=& e^{iKX}\frac{1+(K^2-1)\cosh^2X+2iK \sinh X \cosh X}{(K-i)^2\cosh^2X}
\ \ \ \ \ \ \\
V_k(x) & =& \frac{e^{iKX}}{(K-i)^2\cosh^2X}
\eea
where (differently from section \ref{sec:asympt}) 
we have set 
$X=x/(2\xi)$ and $K=2k\xi$.
The modes for $k<0$ are deduced from the relations $U_k(x)=U_{-k}(-x)$
and $V_k(x)=V_{-k}(-x)$.

In presence of the trap, there is an extra contribution to the Hamiltonian
containing the trapping energy,
\be
\mathcal{W} = \frac{1}{2} m\omega^2(t)  \int_{\RR} dx\, x^2 \psi^*(x)
\psi(x).
\ee
We approximate it to leading non-trivial order in $\Lambda$,  that is to first 
order,
injecting the splitting (\ref{eq:splitting}). The term in $a_0^* a_0$
deviates from  $N$ by $O(\Lambda^2)$ so it gives to first order
only a constant contribution, with an integral over $x$
that can be calculated exactly if necessary. 
The crossed terms $a_0^*\Lambda$ and complex
conjugate are kept, with $a_0$ approximated by $N^{1/2} e^{i\theta}$,
whereas the quadratic terms $\Lambda^* \Lambda$ are neglected.
Finally, we replace the field $\Lambda(x)$ by its modal expansion
\begin{multline}
\Lambda(x) = -N^{1/2} \phi_0'(x)\, Q 
+\frac{i}{\hbar N^{1/2}}\, x\, \phi_0(x)\, P \\
+\int_{\RR} \frac{dk}{2\pi} \left[
u_k(x) b_k + v_k^*(x) b_k^*\right].
\end{multline}
We thus keep
\be
\mathcal{W} \simeq \frac{1}{2} N m\omega^2(t) \int_{\RR} dx\, x^2 \phi_0^2(x)
+\chi(t)\, \int_{\RR} \frac{dk}{2\pi} 
\left[
\gamma_k b_k + \gamma_k^* b_k^*
\right],
\ee
with coefficients
\be
\gamma_k = N^{1/2} \frac{1}{2} m\omega^2 \int_{\RR} dx\, 
x^2 \phi_0(x) [u_k(x) + v_k(x)].
\ee
With the residues method, the resulting integral 
can be evaluated analytically. We give for convenience the ratio
to the mode eigenenergy:
\be
\frac{\gamma_k}{\epsilon_k} 
= -\left(\frac{\hbar\omega}{\mu_0}\right)^2
\frac{\pi(N\xi)^{1/2}/2}{(K-i)^2 (1+K^2)}
\frac{1}{\cosh(K\pi/2)}
\label{eq:g_sur_e}
\ee
where $K=2 k\xi$ as above.

When the perturbation $\mathcal{W}$ is added to the unperturbed Hamiltonian
(\ref{eq:normal}), the equilibrium value of the $b_k$'s minimizing
the resulting energy is
\be
b_k(0) = - \frac{\gamma_k^*}{\epsilon_k}.
\label{eq:initial}
\ee
At later times, the trap is opened according to the switch-off function 
$\chi(t)$. The Hamiltonian equations of motion then give
\be
i\hbar \frac{d}{dt} b_k(t) = \epsilon_k b_k(t)  + \chi(t) \gamma_k^*,
\ee
to be solved with the initial conditions  (\ref{eq:initial}):
\be
b_k(t) = b_k(0) \left[\chi(t) -e^{-i\omega_k t}
\int_0^t d\tau\, e^{i\omega_k\tau} \frac{d\chi(\tau)}{d\tau}\right]
\ee
where $\omega_k=\epsilon_k/\hbar$ is the mode frequency.
This allows to calculate the number of excitations after the trap was
switched off,
using $\chi(t)\to 0$ for $t\to +\infty$:
\be
N_{\rm exc} \equiv \lim_{t\to +\infty} \int_{\RR} \frac{dk}{2\pi}
b_k^*(t) b_k(t) 
= \int_{\RR} \frac{dk}{2\pi} |b_k(0)|^2 I(\omega_k)
\label{eq:nexc_gen}
\ee
where $I(\Omega)$ is a spectral density of the switch-off procedure
at frequency $\Omega$:
\be
I(\Omega) = \left|\int_{0}^{+\infty} d\tau\, e^{-i\Omega \tau} 
\frac{d\chi(\tau)}{d\tau}\right|^2.
\ee
We recall that (\ref{eq:nexc_gen}) is valid up to leading order
in $N$ (because of the classical field model)
and to leading order in $\omega^2$ (because of the perturbative treatment of
the deviations of the field from the free space soliton).

\begin{figure}[htb]
\includegraphics[width=8cm,clip]{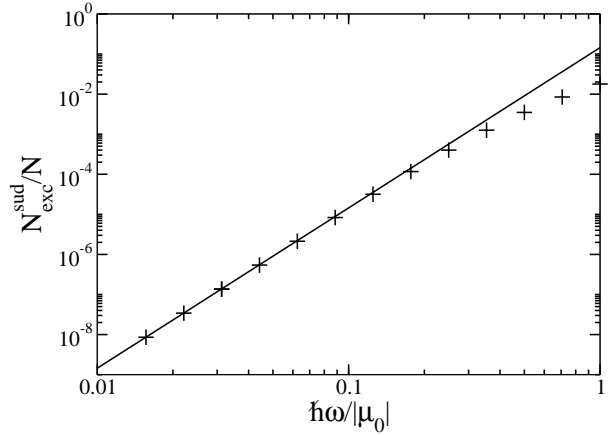}
\caption{For an initially harmonically trapped classical soliton, 
number of excitations $N_{\rm exc}^{\rm sud}$ produced
by a sudden opening of the trap, divided by the number of particles $N$,
and given as a function of $\hbar\omega/|\mu_0|$, in log-log scale.
Solid line: Analytical result (\ref{eq:sud}) obtained in the limit $|\mu_0|\gg\hbar\omega$. 
Symbols: Result deduced from a numerical solution
of the Gross-Pitaevskii equation in the trap.
Here $\omega$ is the oscillation frequency of the particles in the trap, and $\mu_0=-m g^2 N^2/8\hbar^2$
is the chemical potential that the classical soliton would have in the absence of trapping potential,
$m$ being the particle mass and $g$ the coupling constant describing the interactions
in 1D.}
\label{fig:nexc}
\end{figure}

The case producing the maximal number of excitations
for a monotonic $\chi(t)$ 
corresponds to a sudden trap switch-off,
where $I(\Omega)=1$ at all frequencies.
Using (\ref{eq:g_sur_e}), we find that the resulting integral can be evaluated
with the residues method, so that
\be
N_{\rm exc}^{\rm sud} = C N \left(\frac{\hbar\omega}{\mu_0}\right)^4
\label{eq:sud}
\ee
with
\begin{multline}
C 
= \frac{\pi}{16} 
\int_{\RR} dK\, \frac{1}{(1+K^2)^4} \frac{1}{\cosh^2(K\pi/2)}  \\
=\frac{\pi^2 (\pi^2 +25)}{3840}+\frac{\zeta(5) +5\zeta(3)}{128}=
0.1446785\ldots
\end{multline}
This result is encouraging since a moderately small value
$\hbar\omega=|\mu_0|/10$ already leads to a number
of excitations {\sl relative} to the total atom number 
at the $10^{-5}$ level.
As shown in Fig.\ref{fig:nexc} the analytical prediction (\ref{eq:sud}) is in good
agreement with the number of excitations deduced from a
numerical solution of the Gross-Pitaevskii equation,
provided that $|\mu_0|\gg \hbar\omega$.

The number of excitations can be reduced by switching off the trap more slowly.
E.g.\ a linear ramping 
\be
\chi(t) = (1-\lambda t)\, \theta(1-\lambda t)
\ee
leads to
\be
I(\Omega) = 4 \left(\frac{\lambda}{\Omega}\right)^2
\sin^2[\Omega/(2\lambda)].
\ee
Due to the presence of a gap $|\mu_0|$ in the internal excitation spectrum
of the gas, one gets the upper bound on the number of excitations
\be
N_{\rm exc} \leq 
\left(\frac{2\hbar\lambda}{\mu_0}\right)^2 N_{\rm exc}^{\rm sud},
\ee
that is one gains quadratically with the switch-off time when it becomes
longer than the internal soliton time $\hbar/|\mu_0|$.

\section{Conclusion}
\label{sec:conclusion}

Inspired by recent observations of matter wave  bright solitons in atomic
gases,
we have considered here two problems that may be relevant 
for experiments.

The first problem is strictly beyond mean field: It corresponds
to the pair correlations between positions of the particles
in a quantum soliton for a {\sl fixed} position of the center of mass
of the soliton. In particular, we have obtained
analytically the large $N$ limit expression of these pair correlations,
see (\ref{eq:Sasympt}) and (\ref{eq:drho_asympt}).
By integrating (\ref{eq:Sasympt}) over the center of mass position,
we obtain a large $N$ expansion of 
the static structure factor for a fully {\sl delocalized} center of mass position,
a quantity already studied in \cite{Caux1,Caux2}.
On an experimental point of view,
our predictions can be tested
by measuring the positions of the particles in a very broad
quantum soliton, prepared with weak attractive interactions and a relatively
small atom number \cite{condi}.

The second problem was studied in the classical field model.
The number of internal excitations of the gas created by the trap opening
from an initial pure soliton was calculated in the limit where 
the soliton size is smaller than the size of the harmonic single particle
ground state, see (\ref{eq:sud}) for a sudden trap opening.
This also can be seen experimentally by detecting atoms flying
away from the remaining soliton core after the trap opening.
A possible extension of this calculation is to include
quantum fluctuations of the field.

We acknowledge useful discussions with L. Khay\-ko\-vich, C. Weiss,
A. Sinatra, M. Olshanii. Our group is a member of IFRAF.

\appendix

\section{Calculation of the Fourier transforms of $\rho(x|0)$
and $\rho(x,y|0)$}
\label{appen:deriv_gen}

We start with the definition of the mean density for a fixed center of mass
position $R$, taking here $R=0$ without loss of generality: From
(\ref{eq:def_moy}) one has
\begin{multline}
\rho(x|0) = \int_{\RR^N} dx_1\ldots dx_N
\delta\left(\sum_{k=1}^{N} x_k/N\right) \left[\sum_{j=1}^{N}\delta(x_j-x)
\right] \\
\times  |\phi(x_1,\ldots,x_N)|^2.
\end{multline}
Using the bosonic exchange symmetry we can restrict the integral
to the fundamental 
domain $D$ of (\ref{eq:fd}), including a factor $N!$.
For simplicity we take in this appendix $\hbar^2/(m|g|)$ as the unit
of length.
With the change of variables, of Jacobian equal to unity,
\bea
x_j = \sum_{k=1}^{j} u_k, \ \ \ \ 1 \leq j \leq N,
\label{eq:cov}
\eea
the condition to be in $D$ is simply that all the $u_2,\ldots, u_N$ are
positive, and $u_1$ can vary in the whole real space $\RR$.
Setting 
\be
\beta_k=\sum_{j=k}^N [2j-(N+1)]=(N+1-k)(k-1), 
\label{eq:defbk}
\ee
we obtain from (\ref{eq:phi}):
\begin{multline}
\rho(x|0) = [(N-1)!]^2 \sum_{j=1}^{N} \int_{\RR} du_1
\int_{(\RR^+)^{N-1}} du_2\ldots du_N  \\
\delta\left(\sum_{k=1}^{N} \frac{N+1-k}{N} u_k\right) 
\delta\left(x-\sum_{k=1}^{j} u_k\right) e^{-\sum_{k=2}^{N} \beta_k u_k}.
\label{eq:deux_fac}
\end{multline}
One can calculate the integral over $u_1$: The first delta factor in
(\ref{eq:deux_fac}), the one ensuring that $R=0$, imposes
a value
\be
u_1 = - \sum_{k=2}^{N} \frac{(N+1-k)}{N}u_k.
\label{eq:vimp}
\ee
When one reports this value of $u_1$ in the argument of the second delta factor
in (\ref{eq:deux_fac}),
one obtains a remaining factor $\delta(x-\sum_{k=2}^N \alpha_{k,j} u_k)$, 
with
\bea
\alpha_{k,j} &=& \frac{k-1}{N} \ \ \ \ \mbox{for}\  k\leq j \\
         &=& -\frac{N+1-k}{N} \ \ \ \ \mbox{for}\  k > j .
\eea
Taking the Fourier transform of this remaining
delta factor, according to $\tilde{\rho}(q_a|0)=\int_{\RR} dx\ e^{-i q_a x}
\rho(x|0)$, gives
\bea
\tilde{\rho}(q_a|0) &=&  [(N-1)!]^2 \sum_{j=1}^{N} 
\int_{(\RR^+)^{N-1}} du_2\ldots du_N  \nonumber \\
&& \times e^{-iq_a \sum_{k=2}^N \alpha_{k,j} u_k}
e^{-\sum_{k=2}^N \beta_k u_k}  \\
&=& \sum_{j=1}^{N} \prod_{k=2}^{N} \frac{\beta_k}{\beta_k + i q_a \alpha_{k,j}}
\label{eq:fn1}
\eea
where we used the fact that the product of all $\beta_k$ (for $k$ from 2 to $N$) 
is equal to $[(N-1)!]^2$. Replacing the $\alpha_{k,j}$ by their expression and
using the identity
\be
\prod_{k=j}^{N-1} (k+z) = \frac{\Gamma(N+z)}{\Gamma(j+z)}  \ \ \ \ \forall
j \in {1,\ldots,N}
\label{eq:ptg}
\ee
deduced from the basic property $\Gamma(z+1)=z\, \Gamma(z)$ of the 
Gamma function, with the convention that an ``empty" product
is equal to unity, one gets (\ref{eq:rho_fs}).

We now turn to the pair distribution function 
for a center of mass position fixed in $R=0$. Following the same steps 
as for the mean density, we obtain the Fourier transform
\be
\tilde{\rho}(q_a,q_b|0) = \sum_{1\leq j\neq k\leq N} 
\prod_{n=2}^{N} \frac{\beta_n}{\beta_n +
i q_a \alpha_{n,j} + i q_b \alpha_{n,k}}.
\label{eq:fn2}
\ee
We can restrict to a summation over $j<k$, the contribution for $j>k$
being deduced by exchanging $q_a$ and $q_b$.  In the product over $n$,
three ranges have then to be considered, (i) the first range
$n\leq j$, (ii) the mid-range $j+1 \leq n\leq k$, and (iii) the last
range $k+1 \leq n\leq N$. The first range and the last range contributions
can be expressed in terms of the Gamma function as in (\ref{eq:ptg}).
The mid-range contribution is equal to the product
\be
P_{\rm mid} = \prod_{n'=j}^{k-1} \frac{(N-n') n'}{(N-n')n' -i Q_a (N-n') +i Q_b n'}
\label{eq:mid}
\ee
where we have reindexed the product setting $n'=n-1$ and we have defined 
$Q_a=q_a/N$ and $Q_b=q_b/N$. The last step is to consider
the denominator in each factor of (\ref{eq:mid}) as a polynomial
of degree 2 in $n'$: Its roots are $\epsilon_a$ defined in (\ref{eq:trin},
\ref{eq:sol_trin})
and $N+i(Q_a+Q_b)-\epsilon_a$. This leads to
\be
P_{\rm mid} = \prod_{n'=j}^{k-1} \frac{(N-n') n'}{(n'-\epsilon_a)
[N+i(Q_a+Q_b)-\epsilon_a-n']}
\ee
which can now be expressed as a ratio of products of Gamma functions.
Then one gets (\ref{eq:paire_fs}).

\section{Calculation of the variance of $O_2$}
\label{appen:deriv_broad}

We explain how to calculate exactly the moments $\langle O_2^n\rangle_0$
of the quantity
$O_2$ defined in (\ref{eq:defO2}) in the $N$-body internal ground state
for a fixed center of mass position $R=0$.
We take $\hbar^2/m|g|$ as unit of length and we use the transformations
exposed at the beginning of appendix \ref{appen:deriv_gen}. 
Considering the change of variable (\ref{eq:cov}),
we rewrite $O_2$ as
\begin{multline}
O_2 = \sum_{i=1}^N [(x_i-u_1)+u_1]^2 \\
= -N u_1^2 + 2 u_1\left(\sum_{i=1}^{N} x_i \right)+\sum_{i=1}^{N} (x_i-u_1)^2.
\label{eq:2l}
\end{multline}
The first sum in the right hand side of (\ref{eq:2l})
will have a vanishing contribution, 
since the expectation value is taken for a zero center of mass position.
Replacing $x_i-u_1$ by its expression in terms of $u_2,\ldots, u_N$,
and using the fact that the value of $u_1$ is fixed to (\ref{eq:vimp})
we see that one may effectively replace $O_2$ by the quantity
\begin{multline}
O_2 \rightarrow \sum_{i=2}^N \left(\sum_{k=2}^i u_k\right)^2
-\frac{1}{N}\left(\sum_{k=2}^N (N+1-k) u_k\right)^2 \\
\equiv \sum_{i,j=2}^{N} A_{ij} u_i u_j
\end{multline}
with the symmetric matrix
\be
A_{ij} = \frac{1}{N}[N+1-\mbox{max}(i,j)][\mbox{min}(i,j)-1].
\ee
We have thus reduced the problem to the calculation of the integrals
\begin{multline}
\langle O_2^n\rangle_0 = [(N-1)!]^2
\int_{(\RR^+)^{N-1}} du_2\ldots du_N\, \\
\times \left(\sum_{i,j=2}^{N} A_{ij} u_i u_j\right)^n e^{-\sum_{k=2}^N \beta_k u_k},
\end{multline}
with $\beta_k$ defined in (\ref{eq:defbk}).
These integrals may be calculated by interpreting them as
Gaussian averages, introducing the auxiliary complex random variables 
$\alpha_k$, $2\leq k\leq N$: These variables $\alpha_k$
are statistically independent
and each one has a Gaussian probability distribution 
$\propto e^{-|\alpha_k|^2 \beta_k}$.
Each $u_k$ then corresponds to $|\alpha_k|^2$, so that
\be
\langle O_2^n\rangle_0 = 
\langle\langle \left(\sum_{i,j=2}^{N} A_{ij} |\alpha_i|^2 |\alpha_j|^2\right)^n
\rangle\rangle.
\ee
Here $\langle\langle\ldots \rangle\rangle$ denotes the Gaussian average
of the $\alpha_k$'s and can be calculated using Wick's theorem.
The calculations are a bit lengthy for $n=2$ since Wick's theorem
has to be applied to a product of 8 variables. Clearly the matrix
$A_{ij}/(\beta_i \beta_j)$ appears, which is the matrix $B_{ij}$
of (\ref{eq:defB}). We finally get (\ref{eq:sotc})
for $(\mbox{Var}\, O_2)_0= 
\langle O_2^2\rangle_0 -\left(\langle O_2\rangle_0\right)^2$.

\section{Large $N$ asymptotics of $\rho(x,y|0)-\rho(x|0)\rho(y|0)$}
\label{appen:asympt}

To perform the large $N$ expansion we shall use the Fourier
space expressions (\ref{eq:paire_fs},\ref{eq:rho_fs}). We thus define
\be
\delta\tilde{\rho}(q_a,q_b|0) \equiv \tilde{\rho}(q_a,q_b|0)
-\tilde{\rho}(q_a|0) \tilde{\rho}(q_b|0).
\label{eq:def_delta}
\ee
The pair distribution function involves a double sum over $j\neq k$.
Since $\tilde{\rho}(q|0)$ is a simple sum, the last term in (\ref{eq:def_delta})
is a double sum over $j$ and $k$, without the restriction $j\neq k$. It thus
makes sense to split $\tilde{\rho}(q_a|0) \tilde{\rho}(q_b|0)$
 into a off-diagonal part ($j\neq k$), that we
collect with the double sum in $\tilde{\rho}(q_a,q_b|0)$, and a 
diagonal part
\begin{multline}
\mbox{Diag} = \sum_{j=1}^{N} \prod_{\eta=a,b} 
\left|\frac{\Gamma(N)}{\Gamma(N+iQ_\eta)}\right|^2 \\
\times \frac{\Gamma(N+1+iQ_\eta -j) \Gamma(j-iQ_\eta)}{\Gamma(N+1-j)\Gamma(j)},
\end{multline}
where $Q_{a,b}$ are defined above (\ref{eq:trin}).

The idea to obtain the large $N$ limit is simply to replace 
the discrete sums by integrals. To this end, one has to calculate
the large $N$ limit of each term of the sums, for fixed values of
$y_a\equiv j/N$ and $y_b \equiv k/N$.
The expansion of the Gamma functions is conveniently performed using
\be
\frac{\Gamma(z+a)}{\Gamma(z+b)} = e^{(a-b)\ln z}
\left[1+ \frac{(a-b)(a+b-1)}{2z}+ O(1/z^2)
\right]
\label{eq:devut}
\ee
where the real quantity $z$ tends to $+\infty$, and the fixed quantities 
$a$ and $b$ may be complex. One also uses the large $N$ expansion
of the quantity $\epsilon_a$:
\be
\epsilon_a = i Q_a + \frac{Q_a Q_b}{N}+ \ldots
\ee
For the diagonal part, only the leading term of (\ref{eq:devut}) 
is useful. Replacing $\sum_j$ by $N\int dy_a$ leads to
\be
\mbox{Diag} \simeq N \int_0^1 dy_a\, e^{iQ\ln\frac{1-y_a}{y_a}}
=N \int_{\RR} dX_a\, \frac{e^{iQX_a}}{(2\cosh\frac{X_a}{2})^2},
\label{eq:diagl}
\ee
where $Q=Q_a+Q_b$ and the integral was transformed 
with the change of variable $X_a=\ln\frac{x_a}{1-x_a}$, to acquire
the form of a Fourier transform.
Note that the resulting integral can be calculated exactly, giving
$\pi Q/\sinh(\pi Q)$, but this is not useful here.

For the double sum over $j<k$, one has to include the $1/z$ term in the
expansion (\ref{eq:devut}), to obtain a non-zero result:
\begin{multline}
\mbox{Off-Diag} \simeq N Q_a Q_b \int_0^1 dy_a\, \int_{y_a}^1 dy_b\, 
e^{iQ_a \ln\frac{1-y_a}{y_a}}
e^{iQ_b \ln\frac{1-y_b}{y_b}} \\
\times \left[
2+\ln \frac{1-y_a}{y_a} -\ln \frac{1-y_b}{y_b}
-\left(\frac{1}{y_b}+\frac{1}{1-y_a}\right)
\right] + Q_a \leftrightarrow Q_b.
\end{multline}
The change of variables $X_a=\ln\frac{y_a}{1-y_a}$ and $X_b=\ln\frac{y_b}{1-y_b}$
gives to the off-diagonal contribution the form of a Fourier transform:
\begin{multline}
\mbox{Off-Diag} \simeq N Q_a Q_b \int_{\RR^2} dX_a\, dX_b\, 
\frac{e^{-iQ_a X_a} e^{-i Q_b X_b}}{(2\cosh\frac{X_a}{2})^2(2\cosh\frac{X_b}{2})^2}  \\ \times \theta(X_b-X_a) \left[X_b-X_a-e^{X_a}-e^{-X_b}\right]
+Q_a \leftrightarrow Q_b.
\label{eq:offdiagl}
\end{multline}

The leading term of $\delta\tilde{\rho}(q_a,q_b|0)$ for $N\to +\infty$ for 
a fixed $\xi$ is the sum of (\ref{eq:diagl}) and (\ref{eq:offdiagl}).
The Fourier transform with respect to $q_{a,b}$ is straightforward:
The factors $Q_{a,b}$ act as derivatives, and the remaining bits have
already a Fourier form. The Fourier transform of the diagonal contribution
gives a contribution involving a factor $\delta(x_a-x_b)$,
which exactly cancels with the first term in the right-hand side
of (\ref{eq:Svsrho}), at the considered order in $N$, see (\ref{eq:rho_exp}).
We obtain (\ref{eq:Sasympt}).


\begin{thebibliography}{9}
\bibitem{Herzog_a_part}
The calculation was done independently by Christopher Herzog \cite{Herzog},
since the result of \cite{Calogero} was at that time unknown to the authors
of \cite{Herzog}.
\bibitem{aide_sup}
We also used the property $\frac{d}{dX} F(X,X)= 2\partial_X F(X,X) $
where $F(X,Y)$ is the function
over which $\partial_X \partial_Y$ acts in the right
hand side of (\ref{eq:primi}).
\bibitem{details}
One performs the change of variable $Y=X+Z$ in the integral over $Y$,
then one exchanges the order of the integration, integrating first over $X$
(which can be done after the change of variable $u=e^{X}$). One is left with the integral
over $\RR$ of $Z(Z+2)/[\exp(Z)-1]$, which is readily expressed in terms of the Zeta function
after a series expansion of $1/[\exp(Z)-1]$ in powers of $\exp(-Z)$.
\bibitem{lien}
One may wonder, for a fixed $|\mu_0|/(\hbar\omega)$, how large $N$ 
should be to enter the asymptotic regime. In the absence of a detailed
analysis, we give here a naive answer: When one uses Bogoliubov theory
for the quantum field in presence of the trap, one finds that the ground
Bogoliubov mode has an energy exactly equal to $\hbar\omega$,
with mode functions 
$u(x)=[\hbar/(2m\omega)]^{1/2} [(m\omega/\hbar) x\phi(x)-\phi'(x)]$
and $v(x)=[\hbar/(2m\omega)]^{1/2} [-(m\omega/\hbar) x\phi^*(x)-\phi'^*(x)]$,
where $\phi(x)$ is the Gross-Pitaevskii condensate wavefunction normalized
to unity and $\phi'(x)$ is its derivative. This mode corresponds to the center of mass oscillation.
It contributes to the number of non-condensed particles
as $\int_{\RR} dx\, |v(x)|^2$, scaling as $|\mu_0|/(\hbar\omega)$
in the small $\omega$ limit. The naive requirement is thus 
$|\mu_0|/(\hbar\omega) \ll N$.
\bibitem{condi}
Let us assume that the wavefunction $\Phi(R)$ of the center of mass of the gas is a Gaussian wavepacket
of width $\sigma$, centered in $R=0$. To obtain in this case a pair correlation $\rho(x,y)-\rho(x)\rho(y)$
close to the one calculated here for a center of mass perfectly localized in $R=0$,
we estimate that the condition $\sigma< \xi/\sqrt{N}$ should be satisfied, from
the expansion $\rho(x)\simeq \rho(x|0) +\frac{1}{2} \langle R^2\rangle [\partial_R^2\rho(x|R)]_{R=0}$
and $\rho(x,y)\simeq \rho(x,y|0) +\frac{1}{2} \langle R^2\rangle [\partial_R^2\rho(x,y|R)]_{R=0}$,
where $\langle R^2\rangle$ is the expectation value in the wavefunction $\Phi(R)$.
Starting with a harmonically trapped gas in its internal plus center of mass ground state,
as discussed in \cite{Weiss},
one may change the scattering length after trap opening to adiabatically increase the soliton size,
in order to decrease $\sqrt{N}\sigma /\xi$.
\end{thebibliography}
\end{document}